\def\ba{\begin{eqnarray}}
\def\ea{\end{eqnarray}}
\documentclass[12pt]{article}
\usepackage{epsfig}
\begin{document}

\title{Measurement of Renyi Entropies in Multiparticle
Production: a Do-List II}

\author{A. Bialas,
W. Czyz and K.Zalewski \\
M. Smoluchowski Institute of Physics \\
Jagellonian University, Cracow
\thanks{Address: Reymonta 4, 30-059 Krakow,
Poland;
e-mail: bialas@th.if.uj.edu.pl;}\\
Institute of Nuclear Physics, Polish Academy of Sciences,
\thanks{Address: Radzikowskiego 152, 31-342 Krakow, Poland}}

\maketitle

\begin{abstract}
Recently suggested method of measuring  Renyi
entropies of multiparticle systems produced in high-energy collisions is
presented in the form of a ``do-list'', explaining explicitely  how to
perform the measurement and suggesting improvements in the treatment of
the data.
\end{abstract}

PACS{25.75.Gz, 13.65.+i}

\section{Introduction}

A possibility to estimate the Renyi entropies \cite{r} of the
multiparticle systems created in high-energy collisions was suggested
some time ago by two of us \cite{bc2}. The method is based on
observation of the event-by-event fluctuations or, more precisely, on
measurement of coincidences between different events observed in
collisions. Being classical in nature, the analysis of \cite{bc2} could
not, however, provide the absolute scale for entropy and thus the
obtained numbers suffered from a serious uncertainty. Recently, we
published several papers discussing a quantum approach to the same
problem [3--5]. This allows to reduce the
uncertainties of the classical treatment and to formulate the
improvements of the method, leading to a more precise determination of
Renyi entropies from data. Since the Renyi entropy gives the lower limit
on the Shannon entropy \cite{beck}, such measurements may provide
essential information\footnote{For a recent discussion of this point,
see \cite{limit}.} on the structure of the system produced in
high-energy collisions \cite{plasma}.

In the present paper we collect the results of~[3--5] in the form of a ``do
list'', i.e. of an explicit prescription how to perform the measurements and
how to estimate the necessary corrections. We spell out explicitly the steps to
be taken to implement effectively the results of [3--5]. The importance of the
dependence of measurements on discretization of particle momenta and the role
of (multi)particle correlations are emphasized. The paper may be considered as
an extension of the previous publication with a similar title \cite{bc1}, but
it can be read independently.

\section{Selection of the phase-space region}

As the first step in the process of measurement one has to select a
phase-space region in which measurements are to be performed. This of
course depends on the detector acceptance as well as on the physics one
wants to investigate. The region cannot be too large because for large
systems the method is difficult to apply (the requirements on statistics
become too demanding). With a statistics of $10^6$ events, the region
containing (on the average) $\approx 100$ or less particles should be
possible to investigate. A reasonable procedure seems to be to start
from a small region and then increase it until the errors become
unacceptable.

Comment: The proposed measurement is {\it not} restricted to systems with large
number of particles. It can be applied to any multiparticle system, e.g., to
$e^+e^-$ annihilation, hadron--hadron collisions or peripheral nucleus-nucleus
collisions. It was tested for the PYTHIA MC code for $pp$ collisions \cite{fw}
and recently employed for data on hadron--hadron collisions by the NA22
collaboration \cite{kit}.

\section{Discretization of the spectrum}

The selected  region in momentum space\footnote{We use the notation
$K$ for momenta and $P$, $W$ for probabilities.} should now be divided into bins.
 The size of the bins is --- in principle --- arbitrary. It turns out
to be convenient
to express it in the form
\ba
\Delta K_x\Delta K_y\Delta K_z =
\left[\frac{2\pi}{l^{1/(l-1)}}\right]^{3/2}
\frac{\kappa_x}{L_x} \frac{\kappa_y}{L_y}\frac{\kappa_z}{L_z}\,, \label{d1}
\ea
where $l$ is the rank of Renyi entropy to be measured,
$\kappa_x,\kappa_y,\kappa_z$ are arbitrary (positive) parameters, and
 $L_x^2,L_y^2,L_z^2$ are the mean square radii of the system in
configuration space, e.g. \ba L_x^2=\int (X-\bar{X})^2 D(X,Y,Z) dXdYdZ\,.
\label{d2} \ea Here $D(X,Y,Z)$ is the normalized distribution of positions of
particle emission points in configuration space.

As seen from (\ref{d1}) the parameter $\kappa\equiv
\kappa_x\kappa_y\kappa_z$, is a regulator of the size of the bins used
for discretization of the spectrum. Note that $\kappa$ need not be
constant through the selected momentum space. Actually it may be ---
generally --- convenient to vary $\kappa$ with the position of the bin in
momentum space. For example, in case of boost invariant distribution it
is reasonable to discretize with bins of equal size in rapidity. The
well-known equality $dK_z= E(K) dy$ suggests to take
\ba \kappa_z= E(K_z,K_\perp) \kappa_z^{(0)}\,,
\ea
where $E(K_z,K_\perp)$ is the  particle energy in the
considered bin ($K_z$ and $K_\perp$ are the central values in this bin),
 while $\kappa_z^{(0)}$ is a  constant.

Although the value of the measured Renyi entropy does
not depend on the choice of $\kappa$, it should be emphasized that its selection does
influence the accuracy of the measurement.

The number of bins cannot be too large if one wants to keep the
statistical errors under control. It follows that (\ref{d1}) restricts
the acceptable values of $\kappa$ and of the size of the momentum
phase-space region which one may reasonably investigate at a given
statistics of the experiment.

\section{Description of an event}

Using this procedure, an event is characterized by the number of particles in
each bin, i.e. by a set of integer numbers $s\equiv m_j^{(i)}$, where
$j=1,...J$ ($J$ is the total number of bins) and the superscript $(i)$ runs
over all the kinds of particles present in the final state. These sets
represent different states of the multiparticle system which were realized in
the given experiment. The number of possible different sets is, generally, very
large (for 5 bins and 100 indistinguishable particles one obtains about
$5\times 10^6 $ sets). This is, in fact, the main difficulty in the application
of the method. It
 reflects the fact that the system we are dealing with has very
many  possible states.

Comment: It should be realized that, in practice, such a description is never
complete, i.e., it never describes fully the event (even if the bin width is
ignored). Most often some of the variables are summed over. This is the case,
e.g., when one measures only charged particles. Then all the variables (i.e.
multiplicities and momenta) related to neutral particles are summed over. It
may be thus interesting to study {\it reduced events}, when even some of the
measured variables (e.g. particle identity) are summed over (i.e. ignored).

\section{Measurement of experimental coincidence probabilities}

As explained in \cite{bc2}, the measurement of experimental coincidence
probabilities is the basis of the method and therefore the most
important step in the whole procedure\footnote{ The method was adapted
\cite{bcw} to the present context   from the original proposal by Ma
\cite{ma}.}.

The measurement consists of the simple counting how many times ($n_s$)
any given set $s$ appears in the whole sample of events. Once the numbers $n_s$ are known
for all
sets, one forms the sums:
\ba
N{(l)}= \sum_s n_s(n_s-1)...(n_s-l+1)     \label{1}
\ea
with $l=1,2,3,...$\,.  Each sum formally
runs over all the sets $s$ recorded in a given experiment,
but nonvanishing contributions are given
 only by those which were recorded at least $l$ times\footnote{ Since
the number of different sets is very large, most of them shall appear
only once or not at all.}.
 Thus $N(l)$ is the total number of observed coincidences
of $l$ configurations and it can be recognized as the
factorial moments
\cite{bp} of the distribution of $n_s$ (in particular, $N(1)=N$, where
$N$ is the total number of the events in the sample).
The {\it coincidence probability} of  $l$ configurations is thus given by
\ba
C^{\rm exp}(l) = \frac{N(l)}{N(N-1)...(N-l+1)}\approx \frac {N(l)}{N^l}\,.\label{2}
\ea
Of course $C^{\rm exp}(1)=1$. As
explained in \cite{bc2}, this ratio is equal\footnote{The proof follows
closely the argument of \cite {bp}.} to the\break $(l-1)$-th moment of the
probability distribution: $C^{\rm exp}(l) = \sum_s (p_s)^l $.

It is clear that  two identical configurations must have the same
total number of particles measured in the phase-space region considered,
say $M$. It turns out that to obtain Renyi entropies, it is necessary to
determine $C^{\rm exp}(l)$  for each multiplicity separately. We shall denote
these numbers by $C_M^{\rm exp}(l)$.

The error of $C^{\rm exp}(l)$ is determined by the error of the numerator in
(\ref{2}). One finds approximately $[\Delta N(l)]^2 \approx l! N(l)$.

\section{Renyi entropies}

To obtain the Renyi entropies:
\ba
H(l)= \frac1{1-l} \log C(l)  \label{r00}
\ea
 it is necessary to determine the {\it true} coincidence probabilities:
\ba
C(l) = {\rm Tr} [\rho]^l    \label{r0x}
\ea
with $\rho$ being the density matrix of the system.
$C(l)$ can be expressed in terms of the
true coincidence probabilities $C_M(l)$ at fixed multiplicity:
\ba
C(l)= \sum_M [P(M)]^l C_M(l)\,,     \label{r000}
\ea
where $P(M)$ is the multiplicity
distribution\footnote{We remind the reader that  $M$ is
the number of particles taken into account in the measurement.
It need not be identical with the number
of all particles measured  in the part of momentum space considered
 in the analysis.}.

The relation between $C_M(l)$ and the measured $C_M^{\rm exp}(l)$
was studied in [3--5]. It can be summarized as follows:
\ba
C_M(l)=C_M^{\rm exp}(l) {\mit\Lambda}_M (l) {\mit\Psi}_M(l)\,.  \label{re1}
\ea
 The correction factors ${\mit\Lambda}_M$ and ${\mit\Psi}_M$ depend on $M$
(this is the reason why $C_M^{\rm exp}(l)$ must be determined for every
 multiplicity separately). They are discussed in the next two sections.

\section{Estimate of ${\mit\Lambda}_M(l)$}

Denoting the (3M dimensional) normalized momentum
distribution by
\ba
w(K)\equiv e^{-v(K)}, \label{re0}
\ea
and the size of a bin $j$ by $\omega_j$
 the correction factor ${\mit\Lambda}_M(l)$ is given by
 \ba
{\mit\Lambda}_M(l)= \frac{\sum_j \kappa_j^{M} \left\langle  [w(K)]^l \right\rangle _{\omega_j}}
 {\sum_j\kappa_j^{Ml} [\bar{w}_{\omega_j}]^l}\,,
\label{re2}
\ea
where the summation extends over all (3M-dimensional) bins
and $\langle ...\rangle _{\omega_j}$ denotes the average over a bin of volume $(\omega_j)^{M}$,
 e.g.,
\ba \langle [w(K)]^l\rangle _{\omega_j} = \frac1{\omega_j}
\int\limits_{\omega_j} d^{3M}K [w(K)]^l\,. \label{re2a} \ea We have also
introduced the shorthand \ba \bar{w}_\omega(K)\equiv \frac1{\omega}
\int\limits_\omega d^{3M}K w(K)=\langle  w(K)\rangle _\omega . \label{l2} \ea

To estimate ${\mit\Lambda}_M(l)$ we observe that summations in numerator and
denominator of (\ref{re2}) can be expressed as integrals over the considered
phase-space region. We thus have \ba {\mit\Lambda}_M(l)= \frac {\int\limits
d^{3M}K [\kappa(K)]^{M} [w(K)]^l } {\int\limits d^{3M}K [\kappa(K)]^{Ml}
[\bar{w}_\omega(K)]^l }\,.
 \label{l1}
\ea

If $\kappa$ is independent of the bin (i.e. independent of $K$), this formula
simplifies into \ba {\mit\Lambda}_M(l)= \kappa^{-M(l-1)}
\hat{{\mit\Lambda}}_M(l)\;,\qquad \hat{{\mit\Lambda}}_M(l)= \frac {\int\limits
d^{3M}K  [w(K)]^l } {\int\limits d^{3M}K  [\bar{w}_\omega(K)]^l }\,.
\label{l1x} \ea

One sees that in this case $\hat{{\mit\Lambda}}_M(l)$ tends to one
 if the size of the bins is small
enough (i.e. when $w(K)$ can be treated as  constant within one bin). Then the
value of ${\mit\Lambda}_M(l)$ is  under full control.

If the bins are not small enough, one sees from (\ref{l1}) and
(\ref{l1x}) that ${\mit\Lambda}_M(l)$ can be estimated using the MC code
appropriate for the given process. For the numerator this is rather
straightforward. For the denominator, it is necessary to construct first
the ``smeared'' MC which ignores the difference between the momenta of
particles within each bin.

A simpler, but less precise, method is to ignore correlations between
particles and write the distribution $w(K)$ in form of the product
\ba
w(K)= f(K_1)...f(K_M)\,,    \label{ll2}
\ea
where $f(K)$ is the single-particle momentum distribution.

In this case we obtain
\ba
{\mit\Lambda}_M(l)= [\lambda(l)]^M    \label{l3}
\ea
with
\ba
\lambda(l)=  \frac {\int\limits d^{3}K\kappa(K)  [f(K)]^l }
{\int\limits d^{3}K  [\kappa(K)\bar{f}_\omega(K)]^l }
   \label{l4}
\ea
and thus $\lambda(l)$ can be fairly easily evaluated numerically or even
analytically.

\section {Estimate of ${\mit\Psi}_M(l)$}

The second correcting factor ${\mit\Psi}_M$ is given by the formula \cite{bz}
\ba {\mit\Psi}_M(l)= \frac{\int d^{3M} K [w(K)]^l {\mit\Theta}_l (K)} {\int
d^{3M} K [w(K)]^l }\,, \label{re3} \ea where \ba [{\mit\Theta}_l(K)]^{-1}={\rm
Det}\left[1+\sum_{s=1} a_s(l)[T]^s\right] \label{re4} \ea and \ba a_s(l)
=\frac1{2^s}\frac{(l-1)!}{(2s+1)!(l-2s-1)!}\,.    \label{re5} \ea $T$ is the
symmetric $3M\times 3M$ matrix \ba T_{m\alpha,n\beta}
=\frac1{L_\alpha}V_{m\alpha,n\beta} \frac1{L_\beta} \label{re6} \ea with \ba
V_{m\alpha,n\beta}\equiv \frac{\partial}{\partial K_{m\alpha}} \frac{\partial}{
\partial K_{n\beta} } v(K)\,.    \label{re7} \ea Here $m,n =1,...,M$ label the
particles  and $\alpha,\beta =x,y,z$ denote the space directions.

It is seen from these formulae that ${\mit\Psi}_M(l)$ is independent of the
bin size and thus cannot be influenced by selection of $\kappa$. It does
depend, however, on the size of the system in configuration space
($L_x,L_y,L_z$). Moreover, as explained in \cite{bz}, Eq. (\ref{re3})
represents an expansion in powers of $L^{-1}$  and thus can
only be trusted if $L$ is large enough, so that ${\mit\Psi}_M(l)$ is not too
different from one\footnote{It is clear that ${\mit\Psi}_M(l)$ approaches one
in the limit of large $L's$.}.

Note that (\ref{re5}) implies that the sum in (\ref{re4}) is finite,
because all $a_s(l)$ vanish for $s \geq (l-1)/2$. In particular for
$l=2$ we obtain ${\mit\Theta}_2 =1$ and thus also ${\mit\Psi}_M(2) =1$. For the
other two practically interesting cases ($l=3,4$), the sum reduces to
just one term with
\ba
 a_1(3)=\frac1{12}\,,\qquad a_1(4)=\frac1{4}\, . \label{re8}
\ea

Again, a  MC code seems to be the best method to estimate  ${\mit\Psi}_M(l)$.
 Indeed, Eq. (\ref{re3}) can be rewritten as
\ba
{\mit\Psi}_M(l) = \int d^{M} K P_l(K) {\mit\Theta}(K) = \langle {\mit\Theta}\rangle   ,  \label{ps1}
\ea
where the probability distribution $P_l(K)$ is defined as
\ba
P_l(K)=\frac {[w(K)]^l}{\int d^{3}K  [w(K)]^l }\,.   \label{ps2}
\ea
To construct ${\mit\Theta}(K)$, however, an analytic formula for $w(K)$ is
necessary, as  seen from (\ref{re6}) and (\ref{re7}). This may be a
difficulty.

If  correlations between particles are neglected, the matrix
$T_{m\alpha,n\beta}$ is diagonal in $(m,n)$ and the calculation of the
determinant in (\ref{re4}) is greatly simplified.  We write
\ba
 v(K) = \sum_{m=1}^M u(K_m) \label{e1}
 \ea
and thus
\ba
 V_{m\alpha,n\beta}= \delta_{mn}
\partial_{m\alpha}\partial_{m\beta} u(K_m). \label{e2}
\ea
We shall only consider $l=3$ (${\mit\Psi}_M(2)=1$). We have
\ba
[{\mit\Theta}_l(K)]^{-1}={\rm Det}\left[1+\frac1{12}[T]^s\right]= \prod_{m=1}^M
D(K_m), \label{e3}
\ea
 where $D(K_m)$ is the determinant of the $3\times 3 $
matrix
\ba
D(K_m)= {\rm Det}\left[1+\frac1{12}\frac{\partial_{m\alpha}
\partial_{m\beta} u(K_m)}{L_\alpha L_\beta}\right]. \label{e4}
\ea Further simplifications are possible if the system is cylindrically
symmetric, i.e., if $u(K)= u(k_\perp, k_\parallel) $. Then \ba D(K)=
A\left[A^2+A(k_\perp^2 B+\zeta)+k_\perp^2(B\zeta-C^2)\right]\,, \label{e5} \ea
where \ba A&=&1+\frac1{12L_\perp^2}\frac1{k_\perp}\frac{\partial u}{\partial
k_\perp}\,,\quad\;\; B=\frac1{12L_\perp^2 k_\perp^2}\left[\frac{\partial^2
u}{\partial k_\perp^2}-\frac1{k_\perp}\frac{\partial u}{\partial
k_\perp}\right]\,,\nonumber \\
C&=&\frac1{12L_\perp L_\parallel
k_\perp}\frac{\partial^2 u} {\partial k_\parallel \partial
k_\perp}\,,\quad \zeta=\frac1{12L_\parallel^2}\frac{\partial^2
u}{\partial k_\parallel ^2} -\frac1{12L_\perp^2}\frac1{k_\perp}\frac{\partial u}{\partial
k_\perp} \label{e6}
\ea
and the correction is given by
\ba
{\mit\Psi}_M(3)= \left[\frac {\int  \left[e^{-3u(K)}/D(K)\right]
dk_\perp^2 dk_\parallel}
{\int  e^{-3u(K)}dk_\perp^2 dk_\parallel}\right]^M\,. \label{e7}
\ea

\section{Remarks on size of the region in momentum space}

The main difficulty in the measurements is to find a sufficient number of
coincidences (to keep the statistical error under control). Therefore it is
necessary to limit the size of the region in momentum space where the
measurement is performed. This reduces the number of particles and thus
increases the probability of coincidence.  Below we estimate the
practical consequences of this requirement.

\subsection {Consequences of boost-invariance}

Consider first the longitudinal momentum. If the system is approximately
boost-invariant, one expects an approximate linear  relation between the
considered interval in longitudinal momentum ($\delta K_z$) and the size of the
corresponding region in the configuration space ($L_z$, defined as
 the region from which the emitted particles   end up in $\delta K_z$):
\ba
L_z \approx \frac{\delta K_z}{h^2} \,,
\ea
where the proportionality coefficient can be approximated by \cite{g}
\ba
h^2 = \frac{m_\perp}{\tau_0}
\ea
while $m_\perp$ is the transverse mass of the produced particle and
$\tau_0$ is its proper time at freeze-out.

The conclusion is that the size of the selected region in longitudinal
momentum determines the size of the corresponding region in
configuration space. Consequently, the size of the selected region in
longitudinal momentum cannot be too small (if we want the size $L_z$ in
z-direction to be large enough for the analysis of this paper to be
valid). Note that these remarks do not refer to the choice of the
binning (discretization) but to the size of the momentum region in which
the measurement is performed\footnote{By an appropriate selection of
$\kappa_z$, bins can be fixed at will. See the discussion in Section 3.}.

\subsection{Uncorrelated distribution in azimuthal angle}

Great improvement in the feasibility of measurement can be obtained if
correlations between particles are weak and can be neglected. This is
particularly effective if there are no correlations between various
segments of the distribution in azimuthal angle. In this case the
probability to observe a coincidence in the full azimuthal angle ($2\pi$)
equals the square of the probability to observe a coincidence in half of the
full angle ($\pi$). Therefore it is enough to observe the coincidences
separately in two regions of size $\pi$ where the coincidence
probability is much larger. Consequently, one needs much fewer events to
obtain a decent statistics of coincidences (and thus a decent error of
the measurement). The effect can be made even stronger if independence is
observed for smaller regions of azimuthal angle. The procedure requires,
of course, a careful checking with the data.

\section{Numerical estimates}

To obtain an idea about the size of corrections ${\mit\Lambda}_M (l)$ and
${\mit\Psi}_M(l)$ we shall now explicitely evaluate them in a simplified model
where particles are uncorrelated and the single particle momentum
distribution is axially symmetric and boost-invariant. The transverse
momentum distribution is taken in the Boltzmann form. Thus we have
\ba
w(k_\perp,k_z)d^2k_\perp dk_z&=& A e^{-\sqrt{k_\perp^2 +m^2}/T}
d^2k_\perp dy\nonumber\\
&=&\frac{Ae^{-\sqrt{k_\perp^2 +m^2}/T}}
{\sqrt{k_\perp^2+k_z^2+m^2}} d^2k_\perp dk_z\,,  \label{ne1}
\ea
where $A$ is a normalization constant chosen such that \mbox{$\int\!\!
w(k_\perp,k_z)d^2k_\perp dk_z \!=\!1$.}
Consider first ${\mit\Lambda}_M(l)= [\lambda(l)]^M $ with
\ba
\lambda(l)= \frac{\frac1{\omega_\perp\delta k_z}\sum_{\rm bins}
\int\limits_{\omega_\perp}
d^2k_\perp e^{-lm_\perp/T}\int\limits_{\delta k_z}  \frac{dk_z}
{[k_z^2+m_\perp^2]^{l/2}}}
{\frac1{[\omega_\perp\delta k_z]^l}\sum_{\rm bins}
\left[\int\limits_{\omega_\perp}
d^2k_\perp e^{-m_\perp/T}\int\limits_{\delta k_z}  \frac{dk_z}
{\sqrt{k_z^2+m_\perp^2}}\right]^l}\,, \label{num1}
\ea
where
\ba
\delta k_z= \frac{\sqrt{2\pi}}{[\sqrt{l}]^{1/(l-1)}L_z}\;,\qquad
\omega_\perp\equiv \pi \delta k_\perp^2
=\frac{2\pi}{l^{1/(l-1)}L_\perp^2}\,. \label{num2}
\ea

In figure 1 the correction factor
\ba
\delta_2 =\frac{H(2)^{\rm exp}-H(2)}{H(2)^{\rm exp}}   \label{num3}
\ea
is plotted as function of $L_\perp/\kappa_\perp $ (with $\kappa_\perp$
independent of $K$)
for $L_z$ = 1 fm, $\kappa_z=1$
 and two values of $T$. The
longitudinal momentum interval was fixed
by $-0.38$ GeV $\leq k_z \leq$ 0.38 GeV. One sees that for $L_\perp/\kappa_\perp$
greater than 1 fm the correction is small.
%rys.1
\begin{figure}[htb]
\centerline{%
\epsfig{file=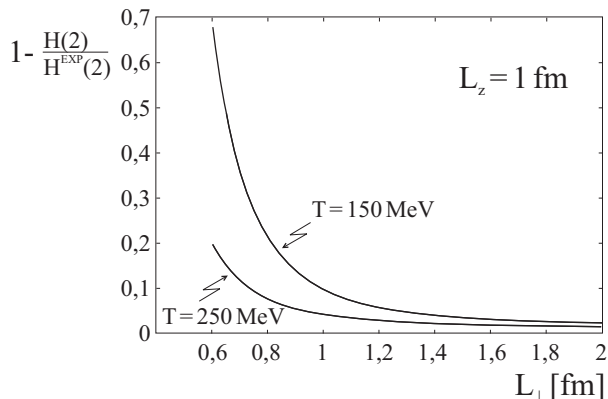,width=8cm}}
\caption{The correction factor $1-H(2)/H^{\rm
exp}(2)$ plotted {\it versus} $L_\perp/\kappa_\perp$ with a constant
$\kappa_\perp$ and $\kappa_z =1$. Other parameters are indicated in the figure.
One sees that for $L_\perp/\kappa_\perp \geq$  1 fm the correction is pretty
small. \label{fig1}}
\end{figure}

It should be emphasized that, as discussed already in Section 7, the
correction factor to $H_2$ can be fully controlled with a good precision
if the bins selected for discretization are small enough. If the size
$L_\perp$ of the system is small, this can be achieved by a proper
choice of the parameter $\kappa$. The results shown in figure 1
demonstrate that to this end it is enough to take $L_\perp/\kappa_\perp$
larger than 1 fm. Since the measurement of $H_2$ provides an effective
lower limit on the value of the entropy of the system \cite{limit}, it
is reassuring that in this simple way the errors can be minimized and
reliably estimated.

For $l\geq 3$ both ${\mit\Lambda}_M(l)=[\lambda(l)]^M$ and
${\mit\Psi}_M(l)\equiv
[p(l)]^M$ are important. This creates a new problem. Indeed, since the
correction factor ${\mit\Psi}_M(l)$ is insensitive to the bin size, it cannot
be eliminated by a proper discretization (${\mit\Psi}_M(l)$ does not depend on
$\kappa$ and thus cannot be adjusted at will).
%rys.2
\begin{figure}[htb]
\centerline{%
\epsfig{file=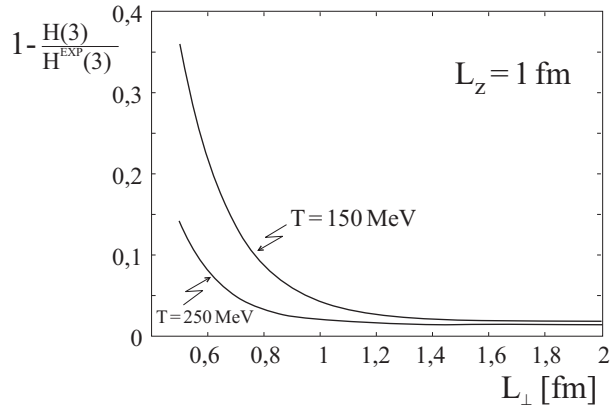,width=8cm}}
\caption{The correction factor $1-H(3)/H^{\rm exp}(3)$ plotted
{\it versus} $L_\perp$ with  $\kappa_\perp=\kappa_z =1$.
\label{fig2}}
\end{figure}

 In figure 2 the correction factor
\ba
\delta_3\equiv \frac{H(3)^{\rm exp}-H(3)}{H(3)^{\rm exp}}  \label{num4}
\ea
is plotted {\it versus} $L_\perp$ for $\kappa_\perp=\kappa_z =1$ and for the
same parameters as in figure 1. It is seen that the corrections are
reasonably small for $L_\perp >  1$ fm but become dangerously large for
smaller radii. Thus for $l\geq 3$ the method seems safe for heavy ion
collisions but cannot be easily justified for systems with linear size
smaller than 1 fm.

\section{Shannon entropy}

The Shannon entropy $S$ (i.e. the standard statistical entropy)
 is formally
equal  to the limit of $H(l)$ as $l\!\!\rightarrow\!\! 1$ and thus can
only be obtained by extrapolation from a
series of measured values: $H(l)=H(2),H(3),...$ to $l=1\,$\footnote{
Obviously, one cannot just put $l=1$ in the formula (\ref{r00})
 for that purpose: since
$C(1) = 1$, the r.h.s. of (\ref{r00}) for $l=1$ represents the undefined symbol
$0/0$. }. Of course such an extrapolation procedure is not unique and
introduces a serious uncertainty \cite{zy}. The main point is, as usual, to
choose the ``best'' extrapolation formula, i.e. the functional dependence of
$H(l)$ on $l$ which will be used to reach the point $l=1$ from the measured
points $l=2,3,...$\,. This form can only be guessed on the basis of physics
arguments (or prejudices).

In \cite{bc2} it was suggested to use
\ba
H(l)= a\frac {\log l}{l-1} + a_0+ a_1(l-1) +a_2(l-1)^2 +\dots\;, \label{5}
\ea
where the number of terms is determined by the number of measured Renyi
entropies. This formula turned out to be very effective in reproducing the
correct value of entropy for some typical distributions encountered in
high-energy collisions.

Another possibility is to use
\ba
H(l) = a_0 + \frac{a_1}{l} + \frac{a_2}{l^2} +\dots  \label{6}
\ea
suggested by the formula for the free gas of massless
bosons\footnote{For the free gas of massless bosons the Renyi entropies
are given by $H(l)=(1+1/l+1/l^2+1/l^3)S/4$ where $S$ is the Shannon
entropy \cite{bc2}.}.

It will be interesting to compare the results from these two formulae.

Comment: The measured values of the Renyi entropies give valuable
information about the system and thus are of great interest,
independently of the accuracy of the extrapolation \cite{bch}.
 Moreover, from the inequality \cite{beck}
\ba
S\geq H(l) \geq H(l+1)\,,   \label{s}
\ea
valid for any $l>1$, we deduce that a
measurement of any Renyi entropy gives an exact lower bound for $S$.
It is well known that this is  important information about the
quark--gluon plasma \cite{plasma}.

\section {Comparison of different regions: Additivity}

Measurements of the  entropies $H(l)$ and  $S$,
 as described above, can be performed independently
(and --- in fact --- simultaneously) in different momentum regions. The results
should give information on the entropy density and its possible dependence on
the region in momentum space (e.g., it seems likely that the results in the
central rapidity region may be rather different from those in the projectile or
target fragmentation region). Furthermore, it is important to verify to what
extent the obtained entropies are additive, i.e., whether the entropies
measured in a region  $R$ which is the sum of two regions $R_1$ and $R_2$
satisfy \ba H(l)(R) =H(l)(R_1)+H(l)(R_2)  \rightarrow S(R)=S(R_1)+S(R_2)\,.
\label{8} \ea
 Eq. (\ref{8}) should be satisfied if there are no strong
correlations between the particles belonging to the regions  $R_1$ and
$R_2$. Thus, verification of (\ref{8}) gives information about the
correlations between different phase-space regions.

Comment: It may be worth  pointing out that the additivity (\ref{8}) can
be more precisely tested for Renyi entropies ($H(l)$) than for the
Shannon entropy ($S$), where the extrapolation procedure (described in
Section 6) always introduces  an additional uncertainty. Since deviations
from additivity signal correlations, this is an interesting problem in
itself.

\section {Conclusions}

In conclusion, we have shown that the measurement of Renyi entropies in
limited regions of phase-space is feasible and thus important
information on the entropy of the system \cite{plasma,limit} is possible
to obtain. Moreover, even the simplest tests of the general scaling and
additivity rules can provide essential information on fluctuations and
on correlations in the system. It should be emphasized that for these
tests the Renyi entropies turn out to be even more useful than the
standard Shannon entropy.

\vspace{0.5cm}

We thank K.~Fia{\l}kowski and W.~Kittel for
discussions.
This investigation was partly supported by the
MEiN research grant 1 P03B 045 29 (2005--2008).

\section *{Appendix:\\Various examples of probability distributions}

\subsection*{Distributions  with exponential tail}

Consider the distribution
\ba
P(r_i)=Ur_i^\alpha e^{-(r_i/a)^\beta}\delta r_i\, , \label{k1}
\ea
where $\delta r_i$ is small, so that we can
replace everywhere the summations by integrals. We also assume that
$\delta r_i = {\rm const} = \delta$. With this assumption the
normalization factor $U$ is given by
\ba
U^{-1} = \int r^\alpha e^{-(r/a)^\beta} dr = \frac1{\beta}a^{(\alpha+1)}
{\mit\Gamma}\left[\frac{\alpha+1}{\beta}\right]\,. \label{k2}
\ea
The distribution (\ref{k1}) covers a wide range of different
distributions. {\it E.g.},  for $(\alpha=0,\,\beta=2)$ one obtains a Gaussian,
for $\beta=1$ (and arbitrary $\alpha$), a Gamma distribution (including, as a special case
the
 exponential distribution).

A fairly straightforward calculation gives
\ba
H(l)= \log\left[\frac{a}{\delta}\frac{{\mit\Gamma}[(\alpha\!+\!1)/\beta]}
{ \beta}\right]\!+\!
\frac1{\beta}\frac{l\alpha\!+\!1}{l-1} \log l  -\frac1{l\!-\!1}
\log\left[\frac{{\mit\Gamma}[(l\alpha\!+\!1)/\beta]}
{{\mit\Gamma}[(\alpha+1)/\beta]}\right]   \label{ak3}
\ea
and thus
\ba
S=\log\left[\frac{a{\mit\Gamma}[(\alpha+1)/\beta]}{\beta\delta}\right]+
\frac{\alpha+1}{\beta}-\frac{\alpha}{\beta} \psi[(\alpha+1)/\beta]\,.
 \label{k4}
\ea
In particular, for a Gaussian we
have\footnote{In this case we take $-\infty \leq r \leq
+\infty$.   The general formulae  (\ref{k1})--(\ref{ak3}) are valid for
 $ 0\leq r\leq +\infty$.}
\ba
H_G(l)=\log\left[\frac{a\sqrt{\pi}}{\delta}\right]+\frac{\log l}{2(l-1)}\,. \label{k3}
\ea

\subsection* {Power law}

\ba
&&\hspace{-5mm}P(r) dr \!=\! \lambda a^\lambda \frac{dr}{(a+r)^{1+\lambda}} \,, \label{ab1}
\\
&&\hspace{-5mm}C(l)\!=\! \lambda^l a^{l\lambda} \delta^{l-1}\!\!
\int\limits_0^\infty\!\!
\frac{dr} {(a\!+\!r)^{l(1+\lambda)}}\!=\!\frac{\lambda^l a^{l\lambda}
\delta^{l-1}}{(l\lambda\!+\!l\!-\!1) a^{l\lambda+l-1}}
\!=\!\left[\frac{\delta}{a}\right]^{l-1}\!\!\! \!\frac{\lambda^l}{l\lambda
+l-1},
\label{ab2}
\ea
\ba
H(l)&=& \log (a/\delta) + \frac1{l-1}
 \log \left(1+(l-1)\frac{1+\lambda}{\lambda}\right) -
\log \lambda\,, \label{ab3}
\\
S&=&\log (a/\delta) +\frac{1+\lambda}{\lambda} - \log \lambda\,.\label{ab4}
\ea

\subsection* {Sum of Gaussians}

Consider probability distribution which is a sum of two identical Gaussians
separated by distance $2R$. This can be written as \ba P(r)
=\frac1{a\sqrt{\pi}} \left[\lambda_- e^{-(r+R)^2/a^2}+ \lambda_+
e^{-(r-R)^2/a^2}\right], \label{a6} \ea where $\lambda_++\lambda_-=1$, and
gives \ba C(l)&=& \frac{\delta^{l-1}}{[a\sqrt{\pi}]^l} \sum_{j=0}^l
\frac{l!}{j!(l-j)!} \lambda_-^j\lambda_+^{l-j}\int dr
e^{-j(r+R)^2/a^2}e^{-(l-j)(r-R)^2/a^2} \nonumber \\&=&
\frac{\delta^{l-1}}{[a\sqrt{\pi}]^{l-1} \sqrt{l}} \sum_{j=0}^l
\frac{l!}{j!(l-j)!}\lambda_-^j\lambda_+^{l-j} e^{-4R^2j(l-j)/(la^2)}\, .
\label{a7} \ea One sees that in the limit of $R/a$ very large, i.e.  for
well-separated Gaussians, only the  terms with $j=0$ and $j=l$  contribute
 and we have
\ba
C(l)= C_G(l) \left(\lambda_-^l+\lambda_+^l\right)\, , \label{a9}
\ea
which simply adds a constant term to the entropy of a single
Gaussian.

If $\lambda_-=\lambda_+=1/2$, we have
$\lambda_-^l+\lambda_+^l=(1/2)^{l-1}$
and thus
\ba H(l) =H_G(l) + \log 2\, . \label{a11}
\ea
It is not difficult to see that for $N$ well-separated
Gaussians one obtains
\ba
 H(l)= H_G(l) -\frac1{l-1} \log
\left(\sum_{i=1}^N (\lambda_i)^l\right) \label{a12}
\ea
If all Gaussians have equal weights $l=1/N$, one obtains
\ba
 H(l)= H_G(l) + \log N  \label{a13}
\ea

This  result is valid for any set of well-separated distributions.

\end{document}